\documentclass[fleqn,twoside]{article}
\usepackage{espcrc2}

\usepackage{natbib}
\bibpunct{[}{]}{}{n}{}{}

\usepackage{graphicx}
\usepackage{amssymb}

\newcommand{\um}[1]{\ensuremath{\,\mathrm{#1}}} 

\newif\ifpdf
\ifx\pdfoutput\undefined
   \pdffalse
\else
   \pdftrue
\fi
\ifpdf
   \newcommand{\ImgPath}[1]{Images/PDF/{#1}.pdf}
\else
   \newcommand{\ImgPath}[1]{Images/EPS/{#1}.eps}
\fi

\title{The AMS-02 Time of Flight System}

\author{D.\ Casadei\address[INFN]{Bologna University and INFN,
        Bologna, Italy}\thanks{Corresponding author: D.\ Casadei,
        Dipartimento di Fisica, Universit\`a di Bologna, via Irnerio
        46, I-40126 Bologna, Italy (\texttt{Diego.Casadei@bo.infn.it}).},
L.\ Baldini\addressmark[INFN],
V.\ Bindi\addressmark[INFN],
N.\ Carota\addressmark[INFN],
G.\ Castellini\address[IROE]{CNR-IROE, Florence, Italy},
F.\ Cindolo\addressmark[INFN],
A.\ Contin\addressmark[INFN],
P.\ Giusti\addressmark[INFN],
G.\ Laurenti\addressmark[INFN],
G.\ Levi\addressmark[INFN],
A.\ Margotti\addressmark[INFN],
R.\ Martelli\addressmark[INFN],
F.\ Palmonari\addressmark[INFN],
L.\ Quadrani\addressmark[INFN],
M.\ Salvadore\addressmark[INFN],
C.\ Sbarra\addressmark[INFN],
A.\ Zichichi\addressmark[INFN]}

\begin{document}

\begin{abstract}
The Time-of-Flight (TOF) system of the AMS detector gives the fast
trigger to the read out electronics and measures velocity, direction
and charge of the crossing particles.  The first version of the
detector (called AMS-01) has flown in 1998 aboard of the shuttle
Discovery for a 10 days test mission, and collected about $10^8$
events.  The new version (called AMS-02) will be installed on the
International Space Station and will operate for at least three years,
collecting roughly $10^{10}$ Cosmic Ray (CR) particles.  The TOF
system of AMS-01 successfully operated during the test mission,
obtaining a time resolution of 120 ps for protons and better for other
CR ions.  The TOF system of AMS-02 will be different due to the strong
fringing magnetic field and weight constraints.

[\emph{Talk given at the ``First International Conference on Particle and
Fundamental Physics in Space'', La Biodola, Isola d'Elba (Italy), 14
-- 19 May 2002.  To be published by Nuclear Physics B - Proceedings
Supplement.}]
\vspace{1pc}
\end{abstract}

\maketitle

\section{Introduction}

The \emph{Alpha Magnetic Spectrometer} (AMS)~\cite{amsfirst} is a
particle detector that will be installed on the International Space
Station in 2005 to measure cosmic ray fluxes for at least three
years.  Amongst the AMS goals we may cite:
\begin{itemize}
  \item Search for cosmic antimatter;

  \item Search for dark matter signatures;

  \item Measurement of primary Cosmic Ray (CR) spectra below 1 TeV:

  \begin{itemize}
    \item Hydrogen and Helium isotopes: solar modulation on a weekly
  basis;

    \item Very high statistics for CR ions below Iron;

    \item Precise measurements of electron and positron spectra;

    \item Cosmic gamma-rays spectrum.
  \end{itemize}
\end{itemize}

\begin{figure*}[t]
\centering
\includegraphics[width=0.7\textwidth]{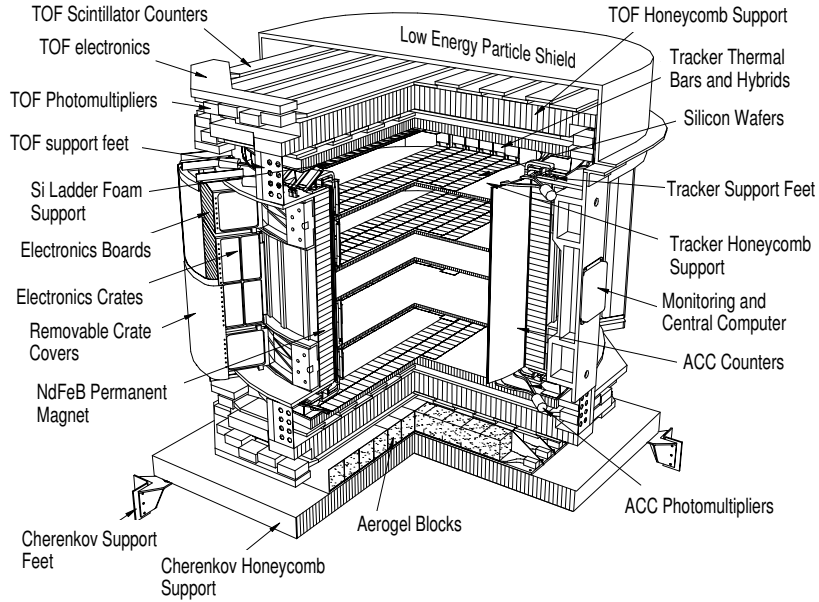}
\caption{The AMS detector for the STS-91 mission (AMS-01).}\label{AMS1}
\end{figure*}

During the precursor flight aboard of the shuttle Discovery (NASA
STS-91 mission, 2--12 June 1998), AMS collected data for about 180
hours~\cite{amsall}.  Figure~\ref{AMS1} shows the test detector
(called AMS-01 in the following), consisting of a permanent Nd-Fe-B
magnet, six silicon tracker planes, a scintillator counters
anticoincidence system, the time of flight (TOF) system consisting in
four layers of scintillator counters and a threshold aerogel
\v{C}erenkov detector.

The TOF system of AMS-01~\cite{tof1} was completely designed and built
at the INFN Laboratories in Bologna, Italy. Its main goals are to
provide the fast trigger to AMS readout electronics, and to measure
the particle velocity ($\beta$), direction, crossing position and
charge.  In addition, it had to operate in space with severe limits
for weight and power consuption (see section~\ref{sec-tof2} for more
details).

\begin{figure*}[t]
\centering
\includegraphics[width=0.7\textwidth]{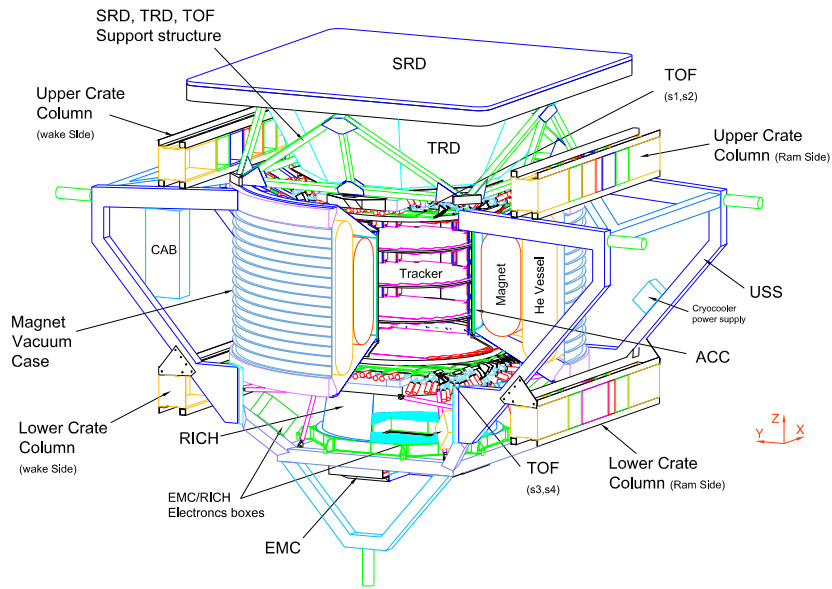}
\caption{The AMS detector to be installed on ISS (AMS-02).}\label{AMS2}
\end{figure*}

Figure \ref{AMS2} shows the new version of the detector (named
AMS-02), that will be installed aboard of the ISS for a 3 years
mission.  AMS-02 will be based on a superconducting magnet that will
produce a dipolar field of 0.85 T maximum intensity (hence the
separating power will be $BL^2 \approx 0.85$ T \um{m^2}).

In addition to a veto system of 8 organic scintillator paddles, a
Silicon tracker of 8 $(x,y)$ planes will be placed inside the magnet.
The new tracking system will reach a spatial resolution of $\sigma_y =
10\, \um{{\mu}m}$ on the bending plane and $\sigma_x = 30\,
\um{{\mu}m}$ for the non-bending plane.  The total active area will be
$\approx 7 \um{m^2}$ ($\sim 2\times10^5$ channels), and its rigidity
resolution will be $\Delta R/R \approx 2\%$ for $R = (1 \div 30)
\um{GV}$ for protons.

The new time of flight system will consist of 4 planes of 8, 8, 10, 8
scintillator counters respectively (see section~\ref{sec-tof2} below).
Its time resolution will be $\sim 140$ ps for protons and better for
higher charged particles, making AMS-02 able to distinguish between
negative and positive charged particles (hence between CR matter and
antimatter) at the $10^{-9}$ level.  The main differences with respect
to the TOF system of AMS-01 are due to the very strong magnetic field
in the photomultiplier tubes (PMT) zone and to more severe weight
limits.

Below the magnet, a proximity focusing RICH will substitute the AMS-01
threshold \v{C}erenkov counter.  With a 2 cm thick aerogel radiator
and a pixel plane 48 cm far from it, by reconstructing the
\v{C}erenkov angle it will reach a velocity resolution
$\Delta\beta/\beta \sim 0.1\%$ and by counting the emitted photons
it will measure the charge of the incident particle.  This instrument
will improve the AMS sensitivity to light elements isotopes up to $(12
\div 13)$ GeV/A and will enhance element discrimination up to Fe.

Finally, two new instruments will be added to the AMS-02 detector with
respect to AMS-01: a Transition Radiation Detector (TRD) on the top,
and an electromagnetic calorimeter (ECAL) at the bottom of the
detector.  By improving the discriminating power amongst electrons and
protons up to high energies, TRD and ECAL will allow AMS-02 to
measure with high statistics the positron and electron spectra up to
$\sim 300$ GeV, covering the most promising energy range for the
search of supersymmetric dark matter particle annihilations.

\section{The AMS-01 TOF}\label{sec-tof1}

\begin{figure*}[t]
\centering
\includegraphics[width=0.7\textwidth]{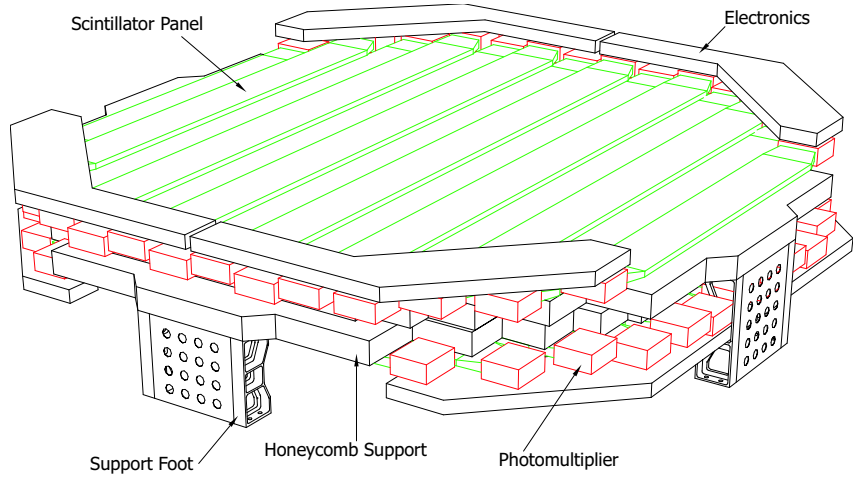}
\caption{Upper two planes of the TOF system of AMS-01.}\label{tof-1}
\end{figure*}

The time of flight of the AMS-01 detector consisted of 4 planes of
Bicron BC408 plastic scintillator paddles, two above and two below the
magnet (figure~\ref{tof-1}).  Each TOF plane had 14 scintillator
counters 1 cm thick covering a roughly circular area of 1.6\um{m^2}.
The scintillation light was guided to 3 Hamamatsu R5900
photomultipliers per side, whose signals were summed together to have
a good redundancy and light collection efficiency.  The total power
consumption of the system (112 channels, 336 phototubes) was
150\um{W}, while its weight (support structure included) was
250\um{kg}.

\begin{figure*}[t]
\centering
\includegraphics[width=0.7\textwidth]{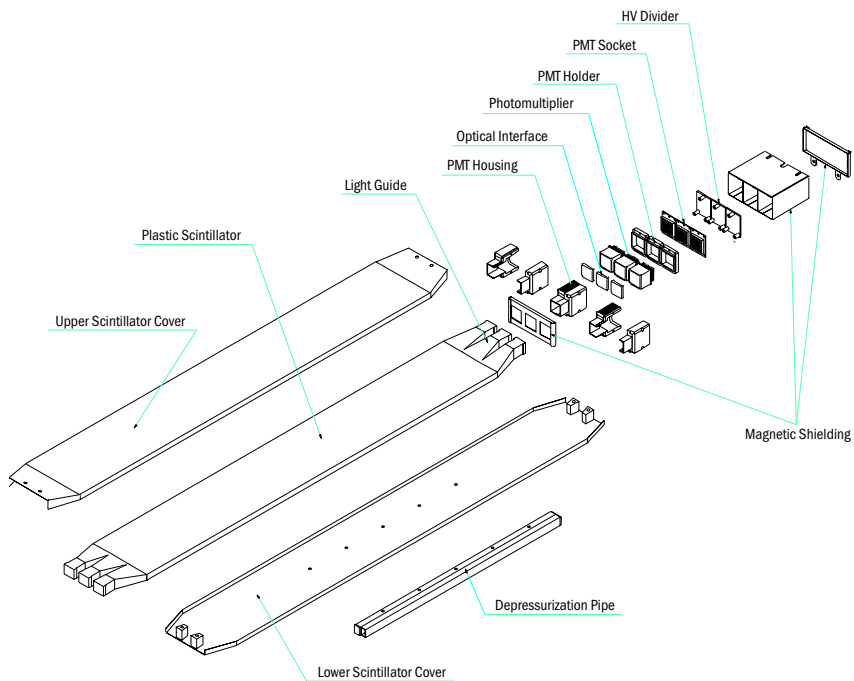}
\caption{Exploded view of a single scintillator counter of the TOF
system of AMS-01.}\label{tof1-cnt}
\end{figure*}

Figure \ref{tof1-cnt} shows the exploded view of a counter: the
scintillator paddle is connected to its Hamamatsu R5900 PMTs through
small plexyglass light guides glued to the counter and soft
transparent optical pads that provide also the mechanical coupling.
The paddle and light guides are surrounded by a thin reflecting mylar
foil and enclosed within a 0.5 mm thick carbon fiber rigid shell (that
has a depressurization pipe for the outgassing phase while reaching
the orbiting altitude), and then it is fixed to a honeycomb plane
connected to the magnet structure by 4 support feet.  The 3 PMTs of
each side are shielded by the residual magnetic field of $\sim 200$ G
by Permalloy boxes.

\begin{figure*}[t]
\centering
\includegraphics[width=0.7\textwidth]{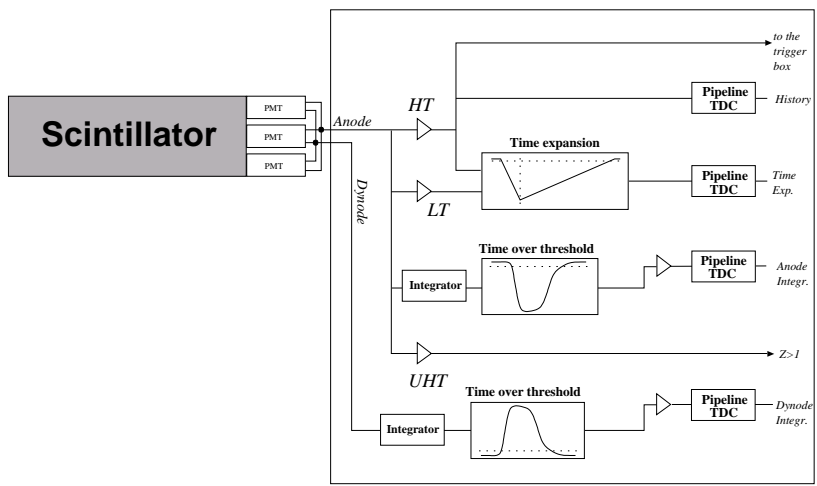}
\caption{Schematic view of the scintillator read-out of the TOF system
of AMS-01.}\label{tof1-ro}
\end{figure*}

Figure \ref{tof1-ro} shows the read-out scheme of a scintillator: the
anode analog signal is split and then is sent to different
discriminators, used to start the ``time expansion'' logic (LT, low
threshold), to send a signal to the trigger box (HT, high threshold)
for the ``fast trigger'' (FT) generation (that is the reference time
of AMS), and to flag He and other nuclei (UHT, ultra-high threshold).
In addition, the anode and dynode signals are used to charge two
capacitors whose discarging times are measured with a pipeline
multichannel TDC (these times are logarithmic functions of the
charge).  

For every counter edge, four digitized data buffers are recorded: the
anode and dynode charges, the time expansion channel (equal to $40
\times$ time between the PMT signal and the FT arrival), and the
``history'' channel (corresponding to the HT), that is used to check
for multiple particle crossing in a 16 \um{{\mu}s} time windows around
the FT.  The TDC sampling rate was 1 GHz, being able to measure the
time between the PMT signal and the FT arrival with 25 ps effective
bins.  The electronic noise, that is the lower limit to the time
resolution of the TOF system, was measured with the STS-91 data to be
$(80 \div 90)$ ps~\cite{icrc2001tof}.

\section{The AMS-02 TOF}\label{sec-tof2}

The counters of the new TOF system are similar to those of AMS-01:
they are wrapped by a reflecting mylar foil and enclosed by thin
carbon fiber shells.  The scintillation light is collected by 2 light
guides per side (3 per side on the external counters of the two
outermost planes), and the PMT anode and dynode signals are summed
together.  Howewer, the shape of the light guides and the phototube
model are different from AMS-01, and the two external counters of each
plane have a different shape.

\begin{figure}[t]\centering
\includegraphics[width=\columnwidth]{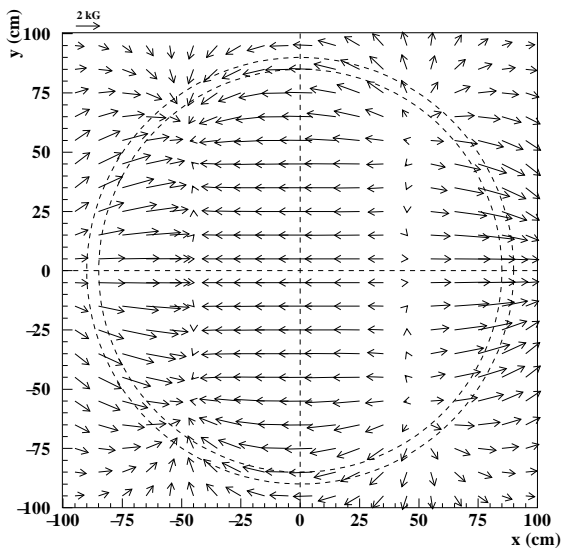}
\caption{Fringing field in the PMT plane of the TOF system of
AMS-02.}\label{tof2-field}
\end{figure}

In fact, the superconducting magnet of AMS-02 is about 5 times
stronger than the permanent magnet of AMS-01, and produces a very
strong field in the zones where the TOF phototubes are positioned.
For this reason, a different kind of PMT has been adopted: the
Hamamatsu R5948 fine mesh model.  In addition, the fringing field has
many different directions (figure~\ref{tof2-field}), making impossible
to adopt straight light guides.  In fact, the measured behavior of the
PMTs depends on the angle between their logintudinal axis and the
magnetic field, making angles wider than $(20 \div 30)$ degrees
unacceptable~\cite{icrc2001pmt}.

In order to minimize the angle between the field direction and the PMT
axis, a variety of bended light guides was produced.  Due to the
combined effect of the choice of the new PMT model and the use of
bended light guides, the expected time resolution is worse than the
AMS-01 one: the new TOF system will achieve a $\approx 140$ ps
resolution for protons, and better for other CR nuclei.

The other strong constraint on the new TOF system design is the
allotted weight budget (240 kg instead of the 250 kg of AMS-01).  In
order to reduce its weight while still keeping at least a 0.4 \um{m^2}
sr geometrical acceptance full covered by the TOF and the tracker, the
number of counters per plane was decreased to 8, 8, 10, 8 (from the
first to last plane, respectively), and the shape of the external
counters was changed.

\begin{figure}[t]\centering
\ifpdf
   \includegraphics[width=\columnwidth]{Images/PDF/tof2-assonom.png}
\else
   \includegraphics[width=\columnwidth]{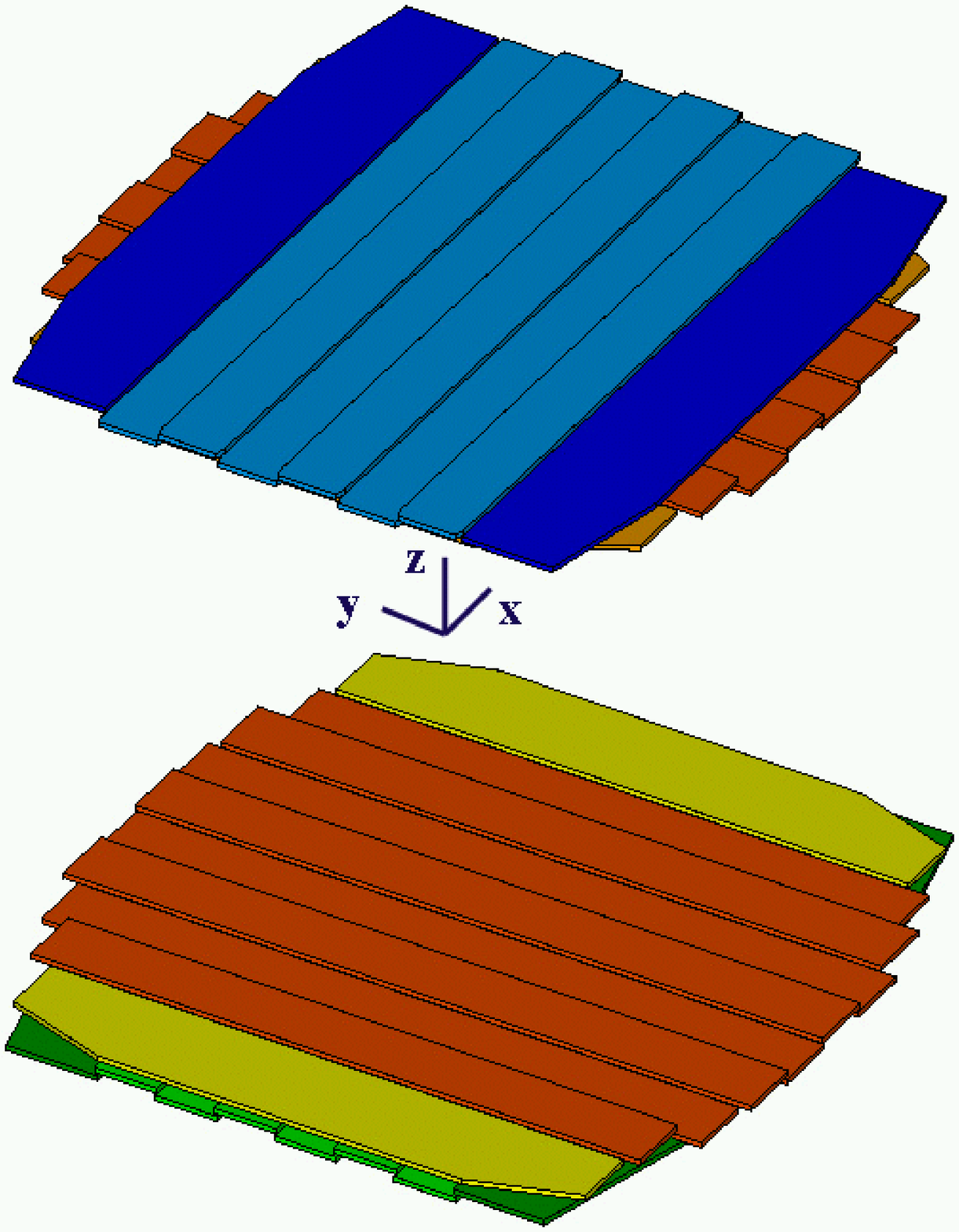}
\fi
\caption{The new layout for the TOF planes of AMS-02.}\label{tof2}
\end{figure}

Figure~\ref{tof2} shows the new TOF planes layout.  The planes 1 and
4, whose counters are parallel to the $x$ axis, have straight and
short light guides and short (130 cm) counters.  These are the planes
with the best time resolution.  The external counters of every plane
have a trapezoidal shape and will be used mainly to check the
consistency of TOF data.

\section{Conclusion}

The time of flight system of AMS-02 will have the same goals of the
TOF system of AMS-01, but it will operate with more severe
conditions.  The low weight and power consumption budgets and the
very strong fringing magnetic field in the phototubes zone will cause
a worsening of the time resolution, with respect to AMS-01.

On the other hand, the new read-out electronics will make the new
system able to reach a better charge resolution than the previous one,
and the trigger efficiency will be unaffected by those problems.  In
addition, fast additional information based on the energy release of the
crossing particles will be sent to the trigger logics, in order to be
able to flag CR nuclei with charge greater than two.



\end{document}